\documentclass{elsart}
\usepackage{amssymb,amsmath,multirow,epsfig,graphicx,color}

\usepackage{epsfig}
\usepackage{graphicx,amsmath}
\newcommand\ba{\begin{eqnarray}}
\newcommand\ea{\end{eqnarray}}

\newcommand{\be}{\begin{equation}}
\newcommand{\ee}{\end{equation}}

\newcommand{\bas}{\begin{eqnarray*}}
\newcommand{\eas}{\end{eqnarray*}}

\newcommand{\cc}{\c c}

\newcommand{\bno}{\begin{eqnarray*}}
\newcommand{\eno}{\end{eqnarray*}}

\def\auj{\number\day\space\ifcase\month\or
janvier\or f\' evrier\or mars\or avril
\or mai\or juin\orjuillet\or ao\^ut
\or septembre\or octobre\or novembre
\or d\' ecembre\fi\space\number\year}

\def\hoje{\number\day\space de \ifcase\month\or
Janeiro,\or Fevereiro,\or Mar\cc o,\or 
Abril,\or Maio,\or Junho,\or Julho,
\or Agosto,\or Setembro,\or Outubro,\or Novembro,
\or Dezembro,\fi\space\number\year}

\def\date{\number\day\space de \ifcase\month\or
January,\or February,\or March,\or 
Appril,\or May,\or June,\or July,
\or Agust,\or septembre,\or Octubre,\or November,
\or December,\fi\space\number\year}

\begin{document}
\begin{frontmatter}

\title{In-Medium Pion Valence Distributions in a Light-Front Model}

\author[UNICSUL]{J. P. B. C de Melo},
\author[UNICSUL]{K. Tsushima}, and
\author[UNICSUL,aff1]{I. Ahmed}
\address[UNICSUL]{Laborat\'orio de F\'\i sica Te\'orica e 
Computacional - LFTC, Universidade Cruzeiro do Sul, 01506-000 S\~ao Paulo, Brazil} 
\address[aff1]{National Center for Physics, Quaidi-i-Azam University Campus, 
Islambad 45320, Pakistan}


\begin{abstract}
Pion valence distributions in nuclear medium and vacuum 
are studied in a light-front constituent quark model. 
The in-medium input for studying the pion properties   
is calculated by the quark-meson coupling model. 
We find that the in-medium pion valence distribution,           
as well as the in-medium pion valence wave function, are substantially  
modified at normal nuclear matter density, due to the reduction in  
the pion decay constant.
\end{abstract}

\end{frontmatter}

{\it Introduction:}~ 
One of the most exciting and challenging topics in hadronic and nuclear physics is 
to study the modifications of hadron properties in nuclear medium (nuclear environment), 
and also how such modifications affect the observables differently from those in vacuum.
Since hadrons are composed of quarks, antiquarks and gluons, it is natural to expect that 
hadron internal structure would change when they are immersed in nuclear medium or 
in atomic nuclei~\cite{Brown,Hatsuda,Saito2007,Hayano,Brooks}.
This question, to study the medium modification of hadron internal structure, 
is particularly interesting when it comes to that of pion. 
To be able to study the properties of pion in nuclear medium, one first needs, simpler, 
effective quark-antiquark models of pion, which are successful in describing  
its properties in vacuum. Among such models, light-front constituent quark model 
has been very successful in describing the hadronic properties in vacuum, in particular, 
the electromagnetic form factors, electromagnetic radii and decay constants 
of pion and kaon~\cite{deMelo1997,deMelo1999,deMelo2002,deMelo2003,daSilva2012,Yabusaki2015,deMelo:2015yxk}.
Recent advances in experiments, indeed suggest to make it possible 
to access to the pion (hadronic) properties in a nuclear 
medium~\cite{Saito2007,Hayano,Brooks,deMelo2014,pimedium2}.

Among the all hadrons, pion is the lightest, and it is believed as 
a Nambu-Goldstone boson, which is realized in nature emerged by  
the spontaneous breaking of chiral symmetry. 
This Nambu-Goldstone boson, pion, plays very important  
and special roles in hadronic and nuclear 
physics~\cite{Fujita:1957zz,Sullivan:1971kd,Coon:1978gr,Ericson:1983um,Weinberg:1991um,Pieper:2001ap,
Frederico2009,Adhikari2016,Fanelli2016aq,Mezrag2015,Chen2016,Hutauruk2016}. 
However, because of its special properties, particularly the unusually light mass,  
it is not easy to describe the pion properties in medium as well as in vacuum  
based on naive quark models, even though such models can be successful in describing 
the other hadrons.

Despite of this difficulty, some important studies were made~\cite{Apion1,Apion2,Apion3} 
on the pion structure and its role in a nuclear medium. 
Recently, we also studied the properties of pion in 
nuclear medium~\cite{deMelo2014,pimedium2}, 
namely, the electromagnetic form factor, charge radius and weak decay constant,  
by using a light-front constituent quark model. There, the in-medium input was calculated  
by the quark-meson coupling (QMC) model~\cite{Saito2007,Guichon}.
We have predicted the in-medium changes of pion properties~\cite{deMelo2014,pimedium2}:    
(i) faster falloff of the pion charge form factor 
as increasing the negative of the four-momentum transfer squared, 
(ii) increasing of the root mean-square charge radius as increasing nuclear density, 
and 
(iii) decreasing of the decay constant as increasing nuclear density.
The purpose of this work is, to extend our work for the pion in medium 
made in Refs.~\cite{deMelo2014,pimedium2}, 
and study the pion valence distribution amplitude in symmetric nuclear matter.
We find substantial modification of the pion valence wave function 
and distribution amplitude in symmetric nuclear matter 
at normal nuclear matter density.

{\it The QMC Model}:~
First, we briefly review the QMC model, 
the quark-based model of nuclear matter, to study the pion properties in medium.
The effective Lagrangian density for a uniform, spin-saturated,
and isospin-symmetric nuclear system (symmetric nuclear matter)
at the hadronic level is given by~\cite{Guichon,QMCfinite}, 
\begin{equation}
{\cal L} = {\bar \psi} [i\gamma 
\cdot \partial -m_N^*({\hat \sigma}) -g_\omega {\hat \omega}^\mu \gamma_\mu ] \psi
+ {\cal L}_\textrm{meson} ~,
\label{lag1}
\end{equation}
where $\psi$, ${\hat \sigma}$ and ${\hat \omega}$ are respectively the nucleon,
Lorentz-scalar-isoscalar $\sigma$, and Lorentz-vector-isoscalar $\omega$ field operators with, 
\begin{equation}
m_N^*({\hat \sigma}) \equiv m_N - g_\sigma({\hat \sigma}) {\hat \sigma}.
\label{effnmass}
\end{equation}
Note that, in symmetric nuclear matter isospin-dependent $\rho$-meson mean filed 
is zero, and thus we have omitted it.
Then the relevant free meson Lagrangian density is given by,
\begin{equation}
{\cal L}_\mathrm{meson} = \frac{1}{2} (\partial_\mu {\hat \sigma} 
\partial^\mu {\hat \sigma} - m_\sigma^2 {\hat \sigma}^2)
- \frac{1}{2} \partial_\mu {\hat \omega}_\nu (\partial^\mu 
{\hat \omega}^\nu - \partial^\nu {\hat \omega}^\mu)
+ \frac{1}{2} m_\omega^2 {\hat \omega}^\mu {\hat \omega}_\mu. 
\label{mlag1}
\end{equation}
Hereafter, we consider the symmetric nuclear matter at rest. 
Then, within Hartree mean-field approximation, 
the nuclear (baryon) and scalar densities are respectively given by,
\begin{eqnarray}
\rho &=& \frac{4}{(2\pi)^3}\int d\vec{k}\ \theta (k_F - |\vec{k}|)
= \frac{2 k_F^3}{3\pi^2},
\label{rhoB}   \nonumber \\ 
\rho_s &=& \frac{4}{(2\pi)^3}\int d\vec{k} \ \theta (k_F - |\vec{k}|)
\frac{m_N^*(\sigma)}{\sqrt{m_N^{* 2}(\sigma)+\vec{k}^2}},
\label{rhos}
\end{eqnarray}
here, $m^*_N(\sigma)$ is the value (constant)  of effective nucleon mass at given density 
(see also Eq.~(\ref{effnmass})).
In the standard QMC model~\cite{Saito2007,Guichon,QMCfinite} the MIT bag model is used, 
and the Dirac equations for the light quarks inside a nucleon (bag) composing nuclear 
matter, are given by, 
\begin{eqnarray}
\left[ i \gamma \cdot \partial_x -
(m_q - V^q_\sigma)
\mp \gamma^0
\left( V^q_\omega +
\frac{1}{2} V^q_\rho
\right) \right]
\left( \begin{array}{c} \psi_u(x)   \\
\psi_{\bar{u}}(x) \\ 
\end{array} \right) &=& 0,
\label{diracu}\\
\left[ i \gamma \cdot \partial_x -
(m_q - V^q_\sigma)
\mp \gamma^0
\left( V^q_\omega -
\frac{1}{2} V^q_\rho
\right) \right]
\left( \begin{array}{c} \psi_d(x) \\   
\psi_{\bar{d}}(x) 
\end{array} \right) &=& 0~.
\label{diracd}
\end{eqnarray}
Because the nuclear matter interactions are strong interactions, the 
Coulomb interaction is neglected as usual, and SU(2) symmetry is assumed, 
~$m_{u,\bar{u}}=m_{d,\bar{d}} \equiv m_{q,\bar{q}}$. 
The corresponding effective (constituent) quark masses are defined  
by, $m^*_{u,\bar{u}}=m^*_{d,\bar{d}}=m^*_{q,\bar{q}} \equiv m_{q,\bar{q}}-V^q_{\sigma}$, 
to be explained later. 

As mentioned already, in symmetric nuclear matter within Hartree approximation, 
the $\rho$-meson mean field is zero, $V^q_{\rho}=0$,~in Eq.~(\ref{diracd}),   
and we ignore it.  The constant mean-field potentials are defined as, 
$V^q_{\sigma} \equiv g^q_{\sigma} \sigma =  g^q_\sigma <\sigma>$, and, 
$V^q_{\omega} \equiv g^q_{\omega} \omega= g^q_\omega <\omega>$,~with $g^q_\sigma$, and 
$g^q_\omega$, are the corresponding quark-meson coupling constants, 
where the quantities with the brackets stand for the expected values 
in symmetric nuclear matter~\cite{Saito2007}. 
Since the average velocity is zero, ~$<\bar{\psi_q} \vec{\gamma} \psi_q> = 0$,
in the nuclear matter rest frame, no spacial-dependent source for the vector-meson 
mean fields arise, and only the terms proportional to~$\gamma^0$ are kept in Eq.~(\ref{diracd}). 
(More details are given in Ref.~\cite{Saito2007}.)

The same meson mean fields $\sigma$ and $\omega$ for the quarks  
in Eqs.~(\ref{diracu}) and~(\ref{diracd}), satisfy self-consistently 
the following equations at the nucleon level:
\begin{eqnarray}
& &{\omega}=\frac{g_\omega \rho}{m_\omega^2},
\label{omgf}\\
& &{\sigma}=\frac{g_\sigma }{m_\sigma^2}C_N({\sigma})
\frac{4}{(2\pi)^3}\int d\vec{k} \ \theta (k_F - |\vec{k}|)
\frac{m_N^*(\sigma)}{\sqrt{m_N^{* 2}(\sigma)+\vec{k}^2}}
=\frac{g_\sigma }{m_\sigma^2}C_N({\sigma}) \rho_s,
\label{sigf}\\
& &C_N(\sigma)=\frac{-1}{g_\sigma(\sigma=0)}
\left[ \frac{\partial m^*_N(\sigma)}{\partial\sigma} \right],
\label{CN}
\end{eqnarray}
where $C_N(\sigma)$ is the constant value of the scalar density ratio~\cite{Saito2007,Guichon,QMCfinite}.
Because of the underlying quark structure of the nucleon used to calculate
$M^*_N(\sigma)$ in nuclear medium (see Eq.~(\ref{effnmass})), 
$C_N(\sigma)$ gets nonlinear $\sigma$-dependence,
whereas the usual point-like nucleon-based model yields unity, $C_N(\sigma) = 1$.

It is this $C_N(\sigma)$ or $g_\sigma (\sigma)$ that gives a novel saturation mechanism
in the QMC model, and contains the important dynamics which originates in the quark structure
of the nucleon. Without an explicit introduction of the nonlinear
couplings of the meson fields in the Lagrangian density at the nucleon and meson level,
the standard QMC model yields the nuclear incompressibility of $K \simeq 280$~MeV with 
$m_q=5$ MeV, which is in contrast to a naive version of quantum hadrodynamics (QHD)~\cite{QHD}
(the point-like nucleon model of nuclear matter),
results in the much larger value, $K \simeq 500$~MeV;
the empirically extracted value falls in the range $K = 200 - 300$ MeV.
(See Ref.~\cite{Stone} for the updated discussions on the incompressibility.)

\begin{table}[t]
\begin{center}
\caption{Coupling constants, and calculated properties for symmetric nuclear matter
at normal nuclear matter density $\rho_0 = 0.15$ fm$^{-3}$,
for $m_q = 5$ and $220$ MeV (the latter values is used in this study and 
was used in Refs.~\cite{deMelo:2015yxk,deMelo2014}). 
The effective nucleon mass, $m_N^*$, and the nuclear
incompressibility, $K$, are quoted in MeV. 
(See Ref.~\cite{Saito2007} for details.)}
\label{Tab:QMC}
\bigskip
\begin{tabular}{c|cccc}
\hline
$m_q$(MeV)&$g_{\sigma}^2/4\pi$&$g_{\omega}^2/4\pi$
&$m_N^*$ &$K$\\
\hline
 5   &5.39 &5.30 &754.6 &279.3  \\
 220 &6.40 &7.57 &698.6 &320.9  \\
\hline
\end{tabular}
\end{center}
\vspace{5ex}
\end{table}

Once the self-consistency equation for the ${\sigma}$ including the quark Dirac equations, 
Eqs.~(\ref{diracu}),~(\ref{diracd}), 
and Eq.~(\ref{sigf}) have been solved, one can evaluate the total energy per nucleon:
\begin{equation}
E^\mathrm{tot}/A=\frac{4}{(2\pi)^3 \rho}\int d\vec{k} \
\theta (k_F - |\vec{k}|) \sqrt{m_N^{* 2}(\sigma)+
\vec{k}^2}+\frac{m_\sigma^2 {\sigma}^2}{2 \rho}+
\frac{g_\omega^2\rho}{2m_\omega^2} .
\label{toten}
\end{equation}
We then determine the coupling constants, $g_{\sigma}$ and $g_{\omega}$, so as
to fit the binding energy of 15.7~MeV at the saturation density $\rho_0$ = 0.15 fm$^{-3}$
($k_F^0$ = 1.305 fm$^{-1}$) for symmetric nuclear matter.

In Refs.~\cite{deMelo:2015yxk,deMelo2014}, the quark mass in vacuum was used 
$m_{q,\bar{q}}~=$~220~MeV to study the pion properties in symmetric nuclear matter.
With this value the model can reproduce the electromagnetic form factor and the decay 
constant well in vacuum~\cite{deMelo2002}.
Thus, we use the same value in this study.
The corresponding coupling constants and some calculated properties for symmetric nuclear matter at 
the saturation density $\rho_0$,
with the standard values of $m_{\sigma}=550$ MeV and $m_{\omega}=783$~MeV, are listed in Table~\ref{Tab:QMC}.
For comparison, we also give the corresponding quantities calculated in the standard QMC
model with a vacuum quark mass of $m_q = 5$~MeV (see Ref.~\cite{Saito2007} for details).
Thus we have obtained the necessary properties of the light-flavor constituent quarks in symmetric
nuclear matter with the empirically accepted data for a vacuum constituent light-quark 
mass of $m_q = 220$~MeV; namely,
the density dependence of the effective mass (scalar potential) and vector potential.
The same in-medium constituent quark
properties which reproduce the nuclear saturation properties 
(and used in Refs.~\cite{deMelo2014,pimedium2}) 
will be used as input to study the pion properties in symmetric nuclear matter.

\begin{figure}[tb]
\begin{center}
\epsfig{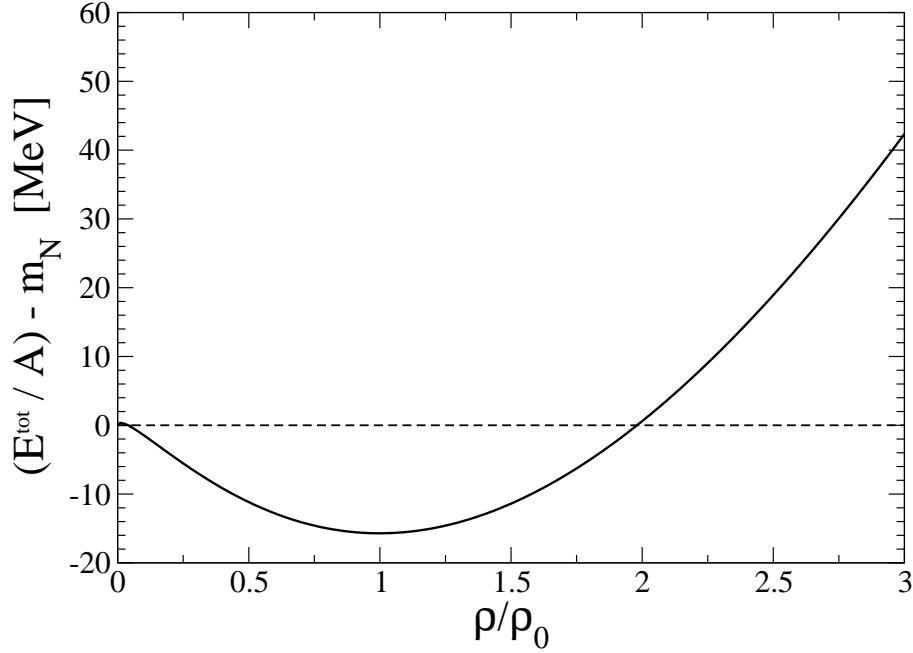}
\caption{Negative of the binding energy per nucleon for symmetric nuclear matter,  
$(E^{tot}/A)-m_N$, as a function of nuclear density $\rho$ ($\rho_0=0.15$ fm$^{-3}$) 
with the vacuum quark mass value $m_q=m_{\bar{q}}=220$~MeV ($q=u,d$), calculated by the 
QMC model (taken from Ref.~\cite{deMelo2014}). 
The corresponding incompressibility $K$ obtained is $K=320.9$ MeV.}
\label{E/A}
\end{center}
\end{figure}

\begin{figure}[tb]
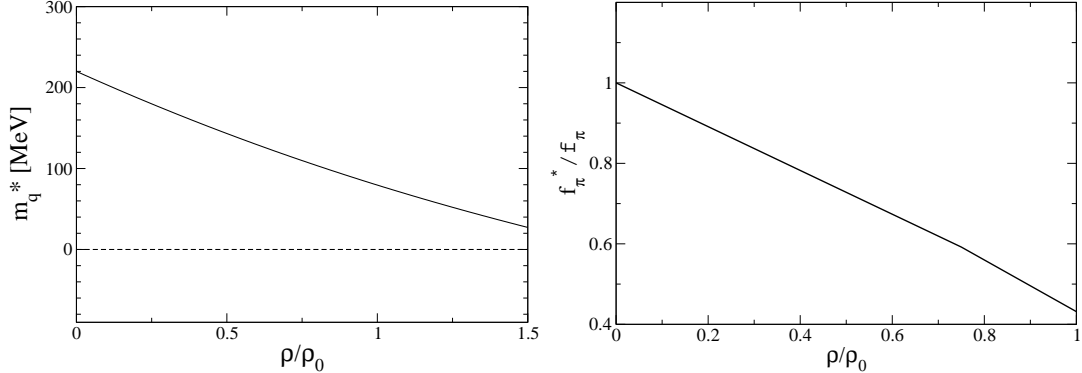

\begin{center}
\hspace*{-1.4cm}
\mbox{
\epsfig{figure=mqstar2.eps,width=7.06cm} 
\epsfig{figure=fpistar2.eps,width=7.06cm} 
} \par
\caption{Effective constituent 
quark mass~$m^*_q$ ($q=\bar{q}=u,\bar{u},d,\bar{d}$)~(left panel), 
and the pion decay constant calculated in symmetric nuclear 
matter~(right panel), both taken from Ref.~\cite{deMelo2014}.
\label{mqfpistar}
}
\end{center} 
\end{figure}

In Figs.~\ref{E/A}~and~\ref{mqfpistar} we respectively show our results for
the negative of the binding energy per nucleon ($E^\mathrm{tot}/A - m_N$),
effective constituent light-quark mass, $m_q^*$, in symmetric nuclear 
matter (left panel of Fig.~\ref{mqfpistar}), 
and the in-medium pion decay constant,~$f^*_{\pi}$ (right panel of Fig.~\ref{mqfpistar}), 
which were calculated in Ref.~\cite{deMelo:2015yxk,deMelo2014}.
For~$f^*_{\pi}$ shown in the right panel of Fig.~\ref{mqfpistar}, 
more explanations will be given later.
Thus, we can say that the in-medium pion properties ($f^*_\pi$ as well), 
are driven by the effective constituent light-quark mass $m^*_q$, 
which is self-consistently calculated and constrained by the symmetric nuclear 
matter saturation properties.

Next, we study the pion valence wave function 
and distribution amplitude~(DA) in symmetric nuclear matter 
using the in-medium constituent light-quark properties obtained so far.

{\it The Model:}~ 
The light-front constituent quark model 
we use here~\cite{deMelo2002,deMelo2003}, although simple, 
is quite successful in describing 
the properties of pion in vacuum, such as the electromagnetic 
form factor, charge radius and weak decay constant. 
The model was also extended for kaon in Ref.~\cite{Yabusaki2015}.
This fact of success in describing properties in vacuum 
is a prerequisite to study the in-medium changes of the pion and kaon properties.
In this study, we focus on the pion.
For some in-medium properties of pion studied in the past, see Ref.~\cite{deMelo2014}.
Note that, we simply use the terminology {\it medium} or {\it nuclear medium} 
hereafter, instead of explicitly specifying {\it symmetric nuclear matter},  
otherwise stated.

To study the in-medium pion properties, we use the input  
calculated by the quark-meson coupling (QMC) model~\cite{Saito2007} 
as mentioned already. 
The QMC model was invented by Guichon~\cite{Guichon} 
to describe the nuclear matter based on the quark degrees 
of freedom. The self-consistent exchange of the scalar-isoscalar $\sigma$ and 
vector-isoscalar $\omega$ mean fields coupled directly to 
the relativistic confined quarks, are the key and novelty for the new  
saturation mechanism of nuclear matter as we explained. The model was extended, and 
has successfully been applied for various nuclear and hadronic phenomena~\cite{Saito2007}.
In the following we briefly summarize the input used for 
the present study of the pion properties in nuclear medium.

The constituent mass of the light quarks ($q$ and ${\bar{q}}$, with $q=u,d$) in the light-front 
constituent quark model in vacuum~\cite{deMelo2014} is, $m_q=m_{\bar{q}}=220$ MeV. 
Then, all the nuclear matter saturation properties are generated by using 
this light-quark mass value.
In other words, the different values of $m_q$ in vacuum generate the corresponding 
different nuclear matter properties, except for the saturation point of the 
symmetric nuclear matter, $\rho = \rho_0$ (normal nuclear matter density, 0.15 fm$^{-3}$) with 
the empirically extracted binding energy of $15.7$ MeV.
This saturation point condition is generally used to constrain 
the models of nuclear matter.

Here, we note that the pion mass up to normal nuclear matter density is
expected to be modified only slightly, where the modification $\delta m_\pi$ 
at nuclear density $\rho = 0.17$ fm$^{-3}$, averaged over the pion isospin states, 
is estimated as $\delta m_\pi \simeq +3$ MeV~\cite{Hayano,Kienle,Meissner,Vogl}.
Therefore, we approximate the effective
pion mass value in nuclear medium to be the same as in vacuum,
$m^*_\pi = m_\pi$, up to $\rho = \rho_0 = 0.15$ fm$^{-3}$, the maximum 
nuclear matter density treated in this study.
Furthermore, since the light-front constituent quark model is rather simple,  
and based on a naive constituent quark picture, the model cannot discuss 
the chiral limit of vanishing (effective) light-quark masses.

We next study the pion properties in symmetric nuclear matter.
The effective interaction Lagrangian density for the 
quarks and pion in medium is given by, 
\begin{eqnarray}
  \mathcal{L}_\mathrm{eff} & = &  -ig^*\, 
  (\bar{q}\gamma^5\vec\tau q \cdot \vec\phi) \ \Lambda^* \ , 
\end{eqnarray}
where the coupling constant,~$g^*=m^*_{q}/f^*_{\pi}$, is obtained by the ''in-medium 
Goldberger-Treiman relation'' at the quark level, with $m^*_{q}$ and $f^*_{\pi}$  
being respectively the effective constituent quark mass 
and pion decay constant in medium, $\vec \phi$ 
the pion field~\cite{deMelo2002,deMelo2003,Yabusaki2015}, and 
$\Lambda^*$ is the $\pi$-$q$-$\bar{q}$ vertex function in medium.
Hereafter, the in-medium quantities are indicated with the asterisk, $^*$.

{\it Symmetric pion valence wave function:}~The pion valence wave function 
used in this study to calculate the pion distribution 
amplitude (poion DA)~\cite{Choi:2007yu,Huang:2013yya},  
(and to be able to calculate also parton distribution function~\cite{Nam2012v1,Nam2012v2}),  
is symmetric under the exchange of quark and antiquark momenta. 
This $\pi$-$q$-$\bar{q}$ vertex function, $\Lambda (k,P)$ in vacuum with 
the arguments $k$ and $P$ stand for momenta, is the same as 
that used for studying the properties of pion~\cite{deMelo2002,deMelo2003,Bennich2008} 
and kaon~\cite{Yabusaki2015}. 
However, for the in-medium $\Lambda^*$, the arguments 
of the function are replaced by those of the in-medium~\cite{deMelo2014}:
\begin{equation}
\Lambda^*(k+V,P)=
\frac{C^*}{((k+V)^2-m^2_{R} + i\epsilon)}+
\frac{C^*}{((P-k-V)^2-m^2_{R}+ i\epsilon)},
\label{vertex}
\end{equation}
where $V^\mu = \delta^\mu_0 V^0$ is the vector potential felt 
by the light quarks in the pion immersed in medium, and can be eliminated by the 
variable change in the $k$-integration, $k^\mu + \delta^\mu_0 V^0 \to k^\mu$.
The normalization factor associated with $C^*$ is modified by the medium 
effects. (See also below Eq.~(\ref{wf2}), and Ref.~\cite{deMelo2014} for details.) 
The regulator mass $m_R$ represents soft effects at short range of about the 1~GeV scale, 
and $m_R$ may also be influenced by in-medium effects. 
However, we employ $m_R^* = m_R$ in Eq.~(\ref{vertex}), 
since there exists no established way of estimating this effect
on the regulator mass. 
This can avoid introducing extra source of uncertainty.

The Bethe-Salpeter amplitude in medium, $\Psi_\pi^*$, with the 
vertex function in medium $\Lambda^*$ is given by,
\begin{eqnarray}
\Psi_\pi^*(k+V,P) & = & \frac{\rlap\slash{k}+\rlap\slash V+m_q^*}{(k+V)^2-m_q^{*2}+ i\epsilon}
\gamma^5 \Lambda^* (k+V,P) \nonumber \\
& & \hspace{25ex} 
\times \frac{\rlap\slash{k}+\rlap\slash V-\rlap\slash{P}+m_q^*}{(k+V-P)^2-m_q^{*2}+ i\epsilon}.
\label{bsa}
\end{eqnarray}
By eliminating the instantaneous terms, namely eliminating 
the terms with gamma matrix $\gamma^+$ in the numerators   
and $k^+$ and~$(P^+ - k^+)$ in the denominators 
with the light-front convention $a^{\pm} \equiv a^0 \pm a^3$, 
and integrating over the light-front energy~$k^-$,  
we obtain the in-medium pion valence wave function $\Phi_\pi^*$,     
\begin{eqnarray}
\Phi_\pi^*(k^+,\vec k_\perp; P^+,\vec P_\perp)=
\frac{P^+}{m^{*2}_\pi-M^2_0} & &  
\left[\frac{N^*}
{(1-x)(m^{*2}_{\pi}-{\cal M}^2(m_q^{*2}, m_R^2))} \right.
\nonumber \\
& & \hspace{3em} \left. +\frac{N^*}
{x(m^{*2}_{\pi}-{\cal M}^2(m^{2}_R, m_q^{*2}))} \right],
\label{wf2}
\end{eqnarray}
where, $N^*=C^* (m^*_{q}/f^*_{\pi}) (N_c)^{\frac{1}{2}}$ 
is the normalization factor with the number of colors $N_c$~\cite{deMelo2002,deMelo2003,deMelo2014}, 
$x=k^+/P^+$ with~$0 \le x \le 1$,  
${\cal M}^2(m^2_a, m_b^2) \equiv  
\frac{\vec{k}^2_\perp+m_a^2}{x}
+\frac{(\vec{P}-\vec{k})^2_\perp+m^2_{b}}{1-x}-\vec{P}^2_\perp \ $,
the square of the mass $M^2_0$ is $M^2_0 ={\cal M}^2(m_q^{*2}, m_q^{*2})$, 
and $m_R$ is the regulator mass with the 
value $m^*_R = m_R = 600$ MeV~\cite{deMelo2002,deMelo2014}. 
Note that the model used in Refs.~\cite{deMelo1999,daSilva2012} 
does not have the second term in Eq.~(\ref{wf2}). 
This means that the pion valence wave function in Refs.~\cite{deMelo1999,daSilva2012} 
is not symmetric under the exchange of quark and antiquark momenta.

The present model with the symmetric 
vertex~\cite{deMelo2002,deMelo2003,Yabusaki2015,Bennich2008}, 
was demonstrated successful in describing the pion properties in nuclear 
medium~\cite{deMelo2014,pimedium2}. 
The pion transverse momentum probability density in medium,   
$P_\pi^*(k_\perp)$, in the pion rest frame $P^+=m_\pi^*$ is 
calculated by, 
\begin{eqnarray}
P_\pi^*(k_\perp)= \frac{1}{4\pi^3 m^*_\pi} \int_0^{2\pi} d\phi 
\int^{m_\pi^*}_0 \frac{d k^{+}M_0^{*2}}
{k^+(m_\pi^*-k^+)} |\Phi_\pi^*(k^+,\vec k_\perp;m^*_\pi,\vec 0)|^2, 
\label{prob1}
\end{eqnarray}
and the integration over $k_\perp$ for $P_\pi^*(k_\perp)$ leads to the in-medium probability 
of the valence component in the pion, $\eta^*$~\cite{deMelo2002,deMelo2003,deMelo2014}:  
\begin{eqnarray}
\eta^*=\int^\infty_0 dk_\perp k_\perp P_\pi^*(k_\perp).
\label{eta}
\end{eqnarray}

The pion decay constant in medium (see Fig.~\ref{mqfpistar} (right panel)), 
in terms of the pion valence component with 
$\Phi_\pi^*(k^+,\vec{k}_\perp;m^*_\pi,\vec{0})$, is calculate by~\cite{deMelo2002,deMelo2014}: 

\begin{eqnarray}
f^*_\pi = \frac{m_q^* (N_c)^{1/2}}{4\pi^3} 
\int \frac{d^{2} k_{\perp} d k^+ }{k^+(m_\pi^*-k^+)} 
\Phi_\pi^*(k^+,\vec{k}_\perp;m^*_\pi,\vec{0}).
\label{fpi}
\end{eqnarray}
%

\begin{table}[t]
\begin{center}
\caption{
Properties of pion in medium, taken from Ref.~\cite{deMelo2014}, with   
$\rho_0=0.15$ fm$^{-3}$.
}
\label{Tab:summary}
\vspace*{3mm}
\begin{tabular}{|c|c|c|c|c|}
\hline
$\rho/\rho_0$  & $m^*_q$~[MeV] & $f^*_{\pi}$~[MeV] & $<r^{*2}_{\pi}>^{1/2}$~[fm] & $\eta^*$ \\
\hline
~0.00  &  ~220    & ~93.1   & ~0.73   & ~0.782   \\
~0.25  &  ~179.9  & ~80.6   & ~0.84   & ~0.812    \\
~0.50  &  ~143.2  & ~68.0   & ~1.00   & ~0.843     \\
~0.75  &  ~109.8  & ~55.1   & ~1.26   & ~0.878      \\
~1.00  &  ~79.5   & ~40.2   & ~1.96   & ~0.930       \\ 
\hline
\end{tabular}
\end{center}
\vspace{8ex}
\end{table}

Some properties of the pion in symmetric nuclear matter 
obtained in Ref.~\cite{deMelo2014}, are summarized in table~\ref{Tab:summary}.
The results listed in table~\ref{Tab:summary} are summarized as follows.
As the nuclear density increases, the in-medium effective constituent quark mass,  
$m_q^*$, and the pion decay constant, $f_\pi^*$, decrease, while the   
root mean square charge radius,  $<r_\pi^{*2}>^{1/2}$, 
and the probability of valence component in the pion state, $\eta^*$, increase.
This can be understood as follows.
The reduction in mass, $m_q^*$, makes it easier 
to excite the valence quark component in the pion, and resulting to 
increase the valence component probability $\eta^*$ in the pion. 
Furthermore, the valence wave function spreads 
more in coordinate space by the decrease of $m_q^*$, 
and reduces the absolute value of the 
wave function at the origin ($f_\pi^* \propto |\Phi_\pi^*(\vec{r}=\vec{0})|$ 
reduction~\cite{Weisskopf}), namely, increases $<r_\pi^{*2}>^{1/2}$.

{\it In-medium pion Distribution Amplitude:} Pion DA provides information 
on the nonperturbative regime of the bound state  
nature of pion due to the quark and antiquark at higher momentum transfer, 
and it was calculated with 
different approaches, such as QCD sum rules~\cite{Mikhailov1986,Bakulev2001}, 
and lattice QCD~\cite{Dalley:2001gj}.
Our study here is based on the light-front constituent quark model.

The pion valence wave function in vacuum is normalized 
by~\cite{Lepage:1982gd,Brodsky:1989pv} 
(aside from the factor $\sqrt{2}$ difference):

\begin{equation}
 \int_0^1 dx \int \frac{d^2 k_\perp}{16 \pi^3 }
 \Phi_{\pi}(x,\vec{k}_\perp)~=~ \frac{f_{\pi}}{2 \sqrt{6}}.
\label{normal}
\end{equation}

This is an important constraint on the normalization of the $q\bar{q}$ wave 
function~\cite{Lepage:1982gd,Brodsky:1989pv}, associated with a probability 
of finding a pure $q\bar{q}$ state in the pion state. 
According to this normalization, the in-medium pion valence wave function 
is normalized by replacing $f_\pi \to f_\pi^*$ in the above.
Since the pion decay constant in nuclear medium is modified, 
the pion valence wave function in nuclear medium is also modified 
via this normalization.

\begin{figure}[tb]
\begin{center}
\hspace*{-1.4cm}
\mbox{
\epsfig{figure=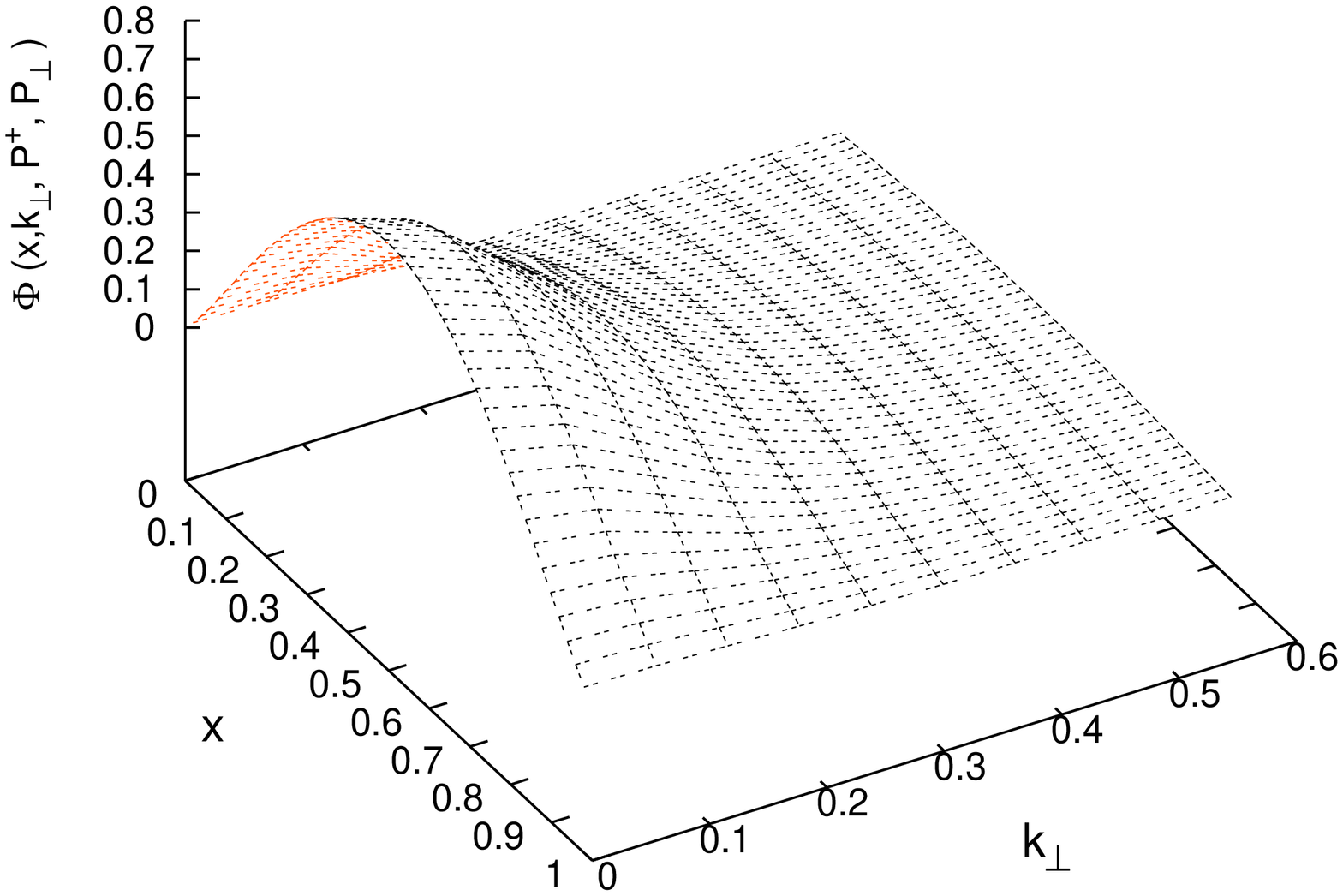,width=6.6cm,angle=-90}
\hspace{-2.3cm}
\epsfig{figure=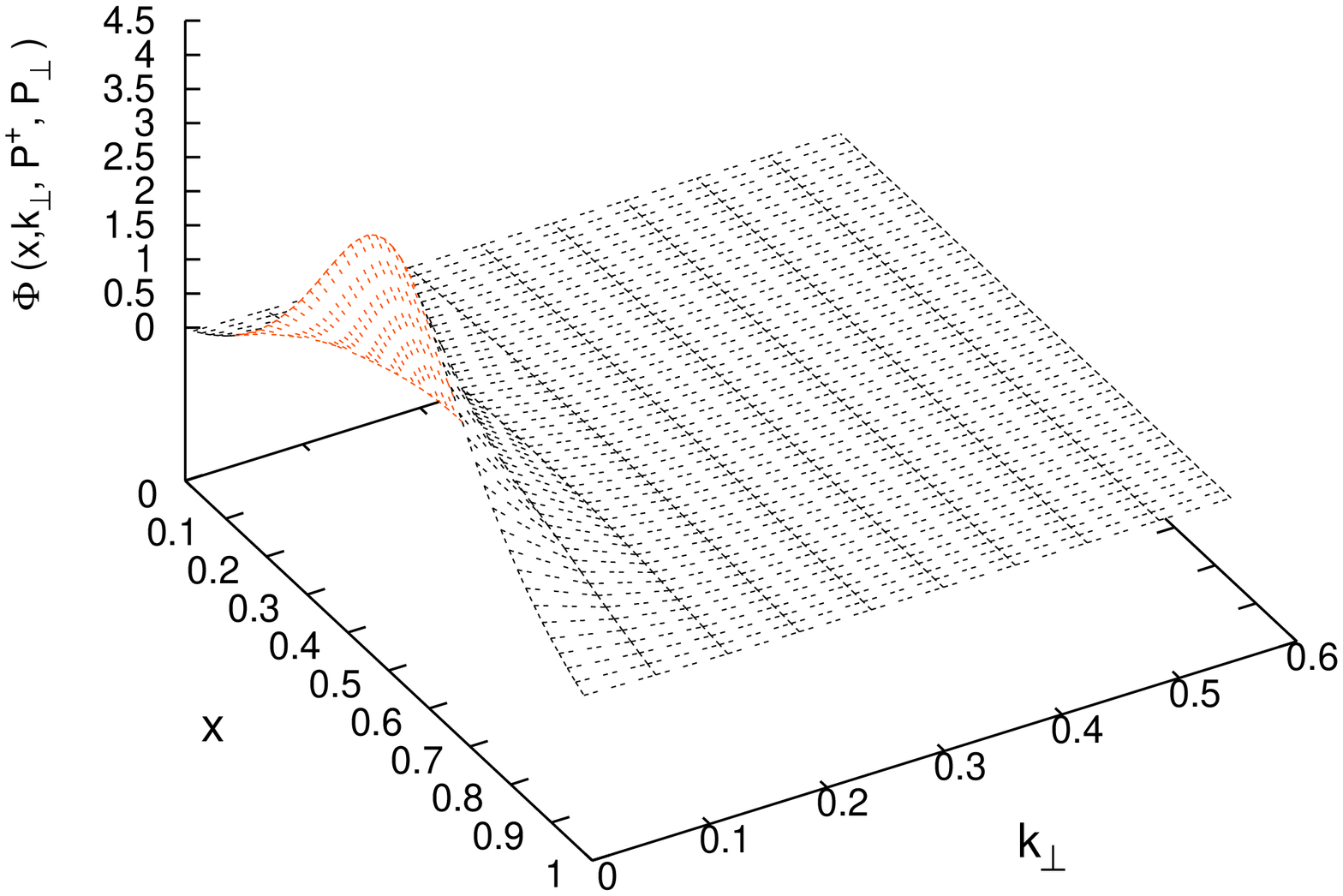,width=6.6cm,angle=-90}
} \par
\caption{Pion valence wave functions in vacuum ($\rho=0$) [left panel] 
and in medium ($\rho=\rho_0$) [right panel]  
v.s. $x$ and $k_\perp = |\vec{k}_\perp|$, where $P^+=m_\pi=m_\pi^*$ and 
$P_\perp=|\vec{P}_\perp|=0$.
The wave functions are given in the units, $10^{-8}\times$(GeV)$^{-1}$. 
Notice that the differences in the vertical axis scales 
for the left and right panels.  
}
\vspace{1.0cm}
\label{wf}
\end{center} 
\end{figure}

In order to examine more in detail as to how the change in $f_\pi^*$ impacts 
on the in-medium pion valence wave function, 
we show in Fig.~\ref{wf} the pion valence wave functions 
in vacuum (left panel) and $\rho=\rho_0$ (right panel).

One can notice that the in-medium pion valence wave function in momentum space 
has a sharper peak and localized in narrower regions both in $x$ and $k_\perp$   
than those in vacuum.
Of course, the total volume, the quantity integrated over $x$ and $\vec{k}_\perp$,  
is reduced in medium, corresponding to the reduced $f_\pi^*$.  
This fact is reflected in the wave function in coordinate space, that 
it becomes spread wider, and generally its hight is reduced.

The corresponding pion valence DA in medium, denoted by $\phi_{DA}^*(x)$  
(not normalized to unity), is calculated as 
\begin{equation}
 \phi_{DA}^*(x)=
 \int \frac{d^2k_{\perp}}{16\pi^3}\Phi_{\pi}^*(x,\vec{k}_\perp).
 \label{Eq:PDA}
\end{equation}
Note that, Eq.~(\ref{Eq:PDA}) holds also for the other pseudoscalar 
mesons $M_{ps}$ such as kaon and D-meson, by replacing 
$\Phi_{\pi}^*(x,\vec{k}_\perp) \to \Phi_{M_{ps}}^*(x,\vec{k}_\perp)$ in the above.

\begin{figure}[tb]
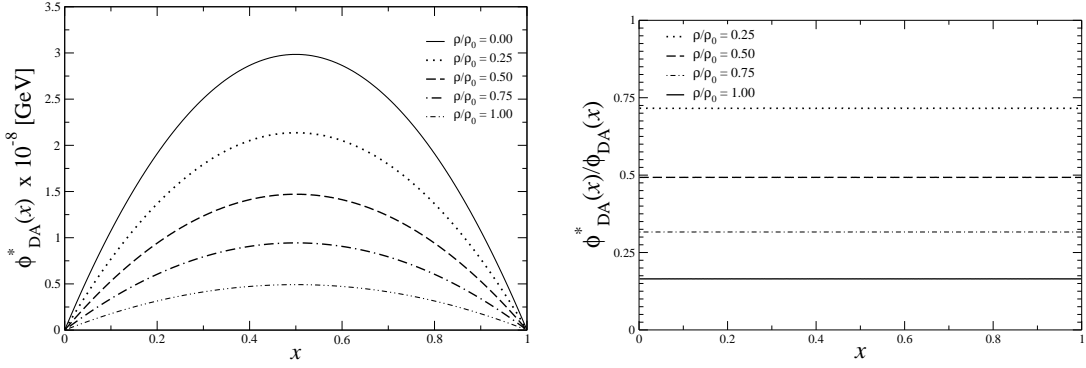

\mbox{
\epsfig{figure=mdapiv1.eps,width=7.0cm}
\hspace{0.3cm}
\epsfig{figure=mdratiosv1.eps,width=6.8cm}
} \par
\caption{Pion valence distribution amplitudes (left panel),   
and the ratios divided by that of the vacuum (right panel).
(See also table.~\ref{Tab:summary}.)}
\label{DA}
\vspace{3ex}
\end{figure}

We show in Fig.~\ref{DA} the pion valence DA, $\phi^*_{DA}(x)$,  
for several nuclear densities including in vacuum $\rho/\rho_0=0$ (left panel),   
and the corresponding ratios divided by the vacuum one $\phi_{DA}(x)$ (right panel). 
Indeed, the significant reduction of the in-medium 
pion valence DAs ($\phi_{DA}^*(x)$) is obvious in Fig.~\ref{DA}, 
reflecting the reduction of $f_\pi^*$.

\begin{figure}[tb]
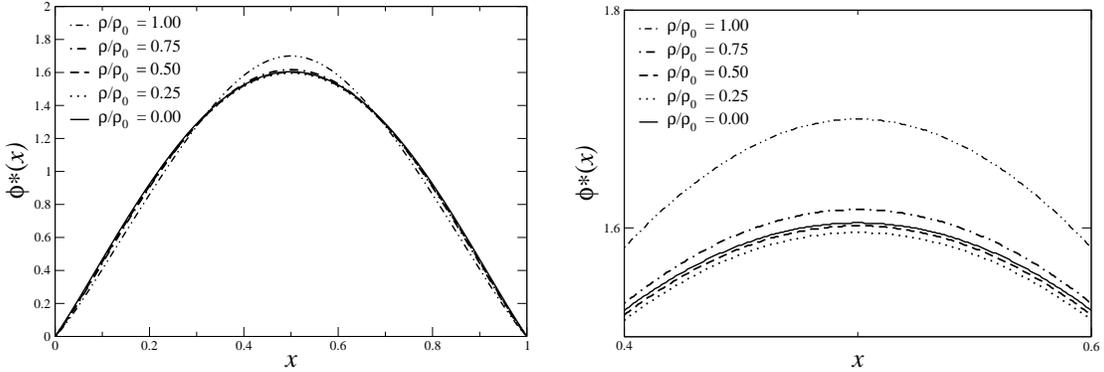

\mbox{
\epsfig{figure=medpdafig1.eps,width=7.0cm}
\hspace{0.3cm}
\epsfig{figure=medpdafig2.eps,width=7.0cm}
} \par
\caption{Normalized pion valence distribution amplitudes (left panel),     
and the magnified ones (right panel), both in vacuum and in medium.
}
\label{NDA}
\vspace{6ex}
\end{figure}

Next, we study pion valence DAs normalized to unity, 
or normalized pion valence DAs in vacuum and in medium. 
By this, we can study the change in shape due to the medium effects.
We show in Fig.~\ref{NDA} the calculated normalized pion valence DAs, 
$\phi^*(x)$ both in vacuum ($\rho/\rho_0=0$) and in medium (left panel), 
and their magnifications (right panel). 
The in-medium change in shape is moderate  
when the nuclear densities are small, 
but it becomes evident when the density becomes $\rho_0$.

\begin{figure}[tb]
\mbox{
\epsfig{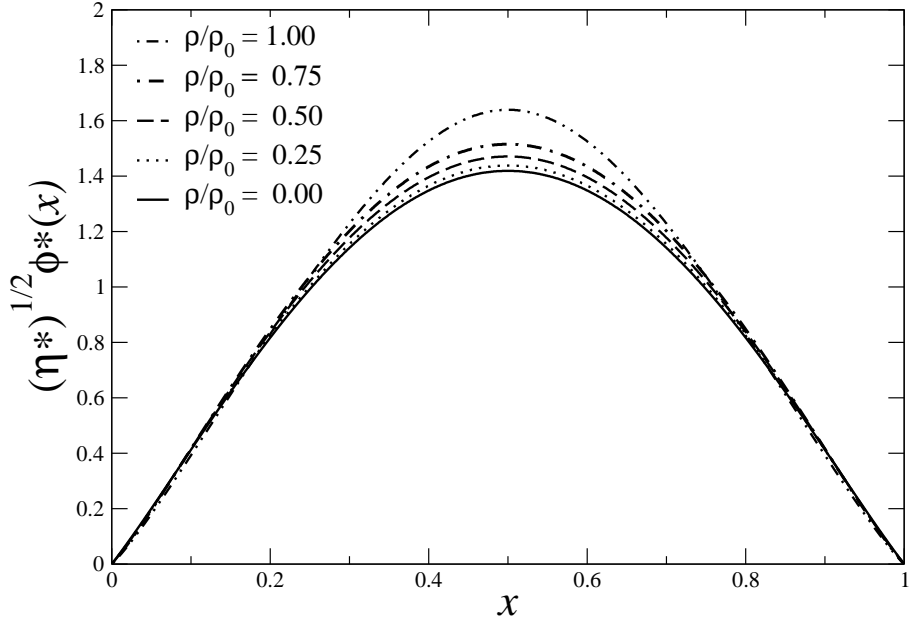}
} \par
\caption{Effective pion valence distribution amplitudes in vacuum and 
in medium, respectively multiplied by $\sqrt{\eta}$ and $\sqrt{\eta^*}$.
}
\label{EFFPDA}
\vspace{3ex}
\end{figure}

Furthermore, it may be useful to define {\it effective pion valence DA} 
using the valence probability in vacuum $\eta$ 
and in medium $\eta^*$. (See Eq.~(\ref{eta}) and table~\ref{Tab:summary}.) 
The pion states in vacuum, $|\pi>$, and in medium, $|\pi>^*$, can respectively 
be written as,
\begin{eqnarray}
|\pi> &=& \sqrt{\eta}|q\bar{q}> + a|q\bar{q}q\bar{q}> + b|q\bar{q}g> + \cdots,
\label{pistate0}\\
|\pi>^* &=& \sqrt{\eta^*}|q\bar{q}>^* + c|q\bar{q}q\bar{q}>^* + d|q\bar{q}g>^* + \cdots,
\label{pistate}
\end{eqnarray}
where $a, b, c$ and $d$  are constants, and $g$ denotes a gluon, 
and $+ \cdots$ stands for the higher Fock components in the pion states.
The quantity $\eta^*$ in table~\ref{Tab:summary} indicates that the valence $q\bar{q}$
component in the pion state increases in medium as nuclear density increases.
The effective pion valence DAs, $\sqrt{\eta^*}\phi(x)^*$, in vacuum ($\rho/\rho_0=0)$ 
and in medium are shown in Fig.~\ref{EFFPDA}. 
They may respectively correspond to the first terms of Eqs.~(\ref{pistate0}) and~(\ref{pistate}).

Since $\eta^*/\eta$ is enhanced in medium, effective pion valence DA   
in medium is also enhanced, on the top of the corresponding 
medium-(shape)modified normalized pion valence DA. 
The obvious enhancement of effective pion valence DA in medium can be seen around $x=0.5$. 
This quantity may be useful when one studies some reactions in medium (in a nucleus)  
involving a pion, based on a constituent quark picture of pion.

{\it Summary:}~We have studied the impact of in-medium effects on the pion valence 
distribution amplitudes using a light-front constituent quark model, 
combined with the in-medium input for the constituent light-quark properties 
calculated by the quark-meson coupling model. 
The in-medium constituent light-quark properties inside the pion are consistently 
constrained by the saturation properties of symmetric nuclear matter. 

The in-medium pion mass is assumed to be the same as that in vacuum, based on the 
extracted information from the pionic-atom experiment, and some theoretical studies. 
This information extracted is valid up to around the normal nuclear matter density. 
Thus, the results obtained in this study, combined with the light-front constituent 
quark model, are valid up to around the normal nuclear matter density, 
but cannot discuss reliably the chiral limit, the vanishing limit of 
the (effective constituent) light-quark masses. We need to rely on more sophisticated models 
of pion to be able to discuss the chiral limit in medium, as well as in vacuum.

Due to the reduction in the pion decay constant 
in medium, the pion distribution amplitude in medium normalized with  
the pion decay constant, is appreciably reduced in nuclear medium.
Because the valence component probability in medium increases as nuclear 
density increases, we have defined an effective pion distribution amplitude 
normalized to the square root of the valence probability in the pion state. 
This may give some information for the effectiveness 
of the valence quark picture of pion in nuclear medium. 
Within the present light-front constituent quark model approach, 
the effectiveness of the valence quark picture of the pion in medium, becomes 
more enhanced as nuclear density increases, due to the increase of the 
valence component in the pion state.

Although the present study is based on a simple, light-front constituent 
quark model, this is a first step to understand the impact of the 
medium effects on the internal structure of the pion immersed in nuclear medium.
In the future, we plan to make similar studies for kaon, D-meson, and $\rho$-meson 
in nuclear medium.

\vspace*{1ex}
{\it Acknowledgements.} 
This work was partly supported by the Funda\c c\~ao de Amparo \`a Pesquisa do Estado de
S\~ao Paulo, Nos. 2015/17234-0,~2015/16295-5,~2016/04191-3
(FAPESP) and Conselho Nacional de Desenvolvimento 
Cient\'ifico e Tecnol\'ogico,~Nos.~400826/2014-3,~401322/2014-9,~308088/2015-8,~308025/2015-6~
(CNPq) of Brazil.

\end{document}